\documentclass[aps,pre,reprint]{revtex4-1}
\usepackage{graphicx}
\usepackage{bm}
\usepackage{amssymb}
\usepackage{epstopdf}

\begin{document}

\title{Tension-dependent transverse buckles and wrinkles in twisted elastic sheets}
\author{Arshad Kudrolli}
\email{akudrolli@clarku.edu}
\affiliation{Department of Physics, Clark University, Worcester, Massachusetts 01610, USA}
\author{Julien Chopin}
\email{julien.chopin@ufba.br}
\affiliation{Instituto de F\'isica,  Universidade Federal da Bahia, Salvador-BA 40170-115, Brazil}

\begin{abstract}
We investigate with experiments the twist induced transverse buckling instabilities of an elastic sheet of length $L$, width $W$, and thickness $t$, that is clamped at two opposite ends while held under a tension $T$. Above a critical tension $T_\lambda$ and critical twist angle $\eta_{tr}$, we find that the sheet buckles with a mode number $n \geq 1$ transverse to the axis of twist. Three distinct buckling regimes characterized as clamp-dominated, bendable, and stiff are identified, by introducing a bendability length $L_B$ and a clamp length $L_{C}(<L_B)$. In the stiff regime ($L>L_B$), we find that mode $n=1$ develops above $\eta_{tr} \equiv \eta_S \sim (t/W) T^{-1/2}$, independent of $L$. In the bendable regime $L_{C}<L<L_B$, $n=1$ as well as $n > 1$ occur above $\eta_{tr} \equiv  \eta_B \sim \sqrt{t/L}T^{-1/4}$. Here, we find the wavelength $\lambda_B \sim \sqrt{Lt}T^{-1/4}$, when $n > 1$. These scalings agree 
with those derived from a covariant form of the F\"oppl-von K\'arm\'an equations, however, we find that the $n=1$ mode also occurs over a surprisingly large range of $L$ in the bendable regime. Finally, in the clamp-dominated regime ($L < L_c$), we find that $\eta_{tr}$ is higher compared to $\eta_B$ due to additional stiffening induced by the clamped boundary conditions.
\end{abstract}

%%%%%%%%%%%%%%% End of first page %%%%%%%%%%%%%%%%%%%%%

\maketitle
\section{Introduction}
Twisting, along with stretching and bending, is a fundamental loading that can be applied to an elastic object. Since the seminal work of Coulomb and Saint-Venant on the elastic equilibrium of prismatic bars~\cite{Coulomb1784,SaintVenant1855}, the response of slender structures under torsion has played a pivotal role in the development of the theory of elasticity~\cite{love2013treatise,timoshenko1953history}. More recently, a large number of studies have focused on complex equilibrium shapes arising upon twisting rods with circular or rectangular crossections due to strong geometrical nonlinearities~\cite{antman1981large,thompson1996helix,Goriely2001,van1998lock,van2003instability}. We focus here on elastic structures such as sheets and ribbons where the thickness is much smaller than the width. In this limit, the classical Kirchhoff theory is ill-suited to model such strongly anisotropic structures and do not accurately predict their torsional stiffness and morphological response~\cite{Dias2015,chopin2016ordered}. Because flexural modes are far less costly energetically than in-plane deformations, thin sheets can undergo buckling instability to accommodate compressive stress~\cite{Audoly2010}. 

While constrained buckling instabilities of elastic rods and sheets in planar configurations have gathered significant attention~\cite{Cerda2003,Chopin2008,Vella2009Macroscopic,Vandeparre2011,brau2011multiple,Davidovitch2011,chopin2017dynamic}, the rich set of buckling and wrinkling patterns observed in elastic sheets upon twist is only being appreciated more recently~\cite{Mockensturm2001,Chopin2013}. Below a critical dimensionless tension $T_{\lambda}$ applied along the longitudinal direction, it is well known that a twisted sheet in the form of a ribbon wrinkles due to the development of compression at the center of the ribbon for sufficient twist~\cite{Green1937,Coman2008}. Twisting above threshold, the wrinkling region is found to widen until reaching the edge of the ribbon consistent with predictions from a recent far-from-threshold theory for ultra-thin sheets~\cite{Davidovitch2011,Chopin2015}. Further, the wrinkling pattern exhibits a symmetry breaking along with a continuous localization of the elastic energy leading to the formation of a triangularly faceted helicoid~\cite{chopin2016disclinations}. The resulting structure called e-helicoid is obtained experimentally under small finite tension as opposed to the developable faceted helicoid which is obtained theoretically for inextensible sheets~\cite{Korte2011}.  However, above $T_{\lambda}$, the ribbon is observed to buckle or wrinkle only in the transverse direction depending on ribbon thickness and length~\cite{Chopin2013,Mockensturm2001,Kit2012}. While the F\"oppl-von K\'arm\'an (FvK) equations commonly used for thin sheets do not give rise to development of destabilizing compression in the transverse direction with twist, an additional term arising from finite rotation effects was identified and included in the instability mechanism~\cite{Chopin2013}. This addition enabled us to capture the thickness dependence of the observed critical twist.

Subsequently, Chopin, Demery, and Davidovitch~\cite{Chopin2015} proposed a covariant extension of the F\"oppl-von K\'arm\'an (cFvK) equations which offers a rigorous theoretical framework to address equilibrium shape of ribbons which  significantly depart from a planar base state. They derived the analytical expression of the transverse and longitudinal stresses, solving perturbatively the cFvK equations using a small slope and small tension limit, and suggested the existence of various transverse instabilities depending on the normalized length $L/W$ and thickness $t/W$, the normalized tension $T$, and the applied boundary conditions.

For a fixed tension $T_{\lambda} \ll T \ll 1$, their model predicts two distinct instabilities of helicoid base state, which was analyzed in depth. They further conjectured a third instability for short ribbons where the base state is dominated by the clamp boundary condition. Previously in Ref.~\cite{Chopin2013}, we showed that the shape of a twisted and stretched ribbon is a helicoid with zero mean curvature and constant negative Gaussian curvature except near the edges. In this region, the clamp boundary condition (a) is not compatible with the helicoid geometry,  and (b) inhibits lateral displacement. Thus, significant shear and transverse stresses are induced on the sheet. Based on energy comparisons between clamped sheet and helicoid ribbon, Chopin, Demery, and Davidovitch~\cite{Chopin2015} also argued that the extent of the deviations from helicoid is given by a clamp length $L_C$ which scales as:
\begin{equation}
    L_{C}=\nu (W^2/t) T^{3/2}.
    \label{Eq:Lc}
\end{equation}
Thus, for $L>L_C$, a twisted sheet is expected to exhibit a helicoid base state. However, the precise distribution of stress inside the clamp-dominated zone is as yet not known.

\begin{figure}
\begin{center}
\includegraphics[width = 8.5cm]{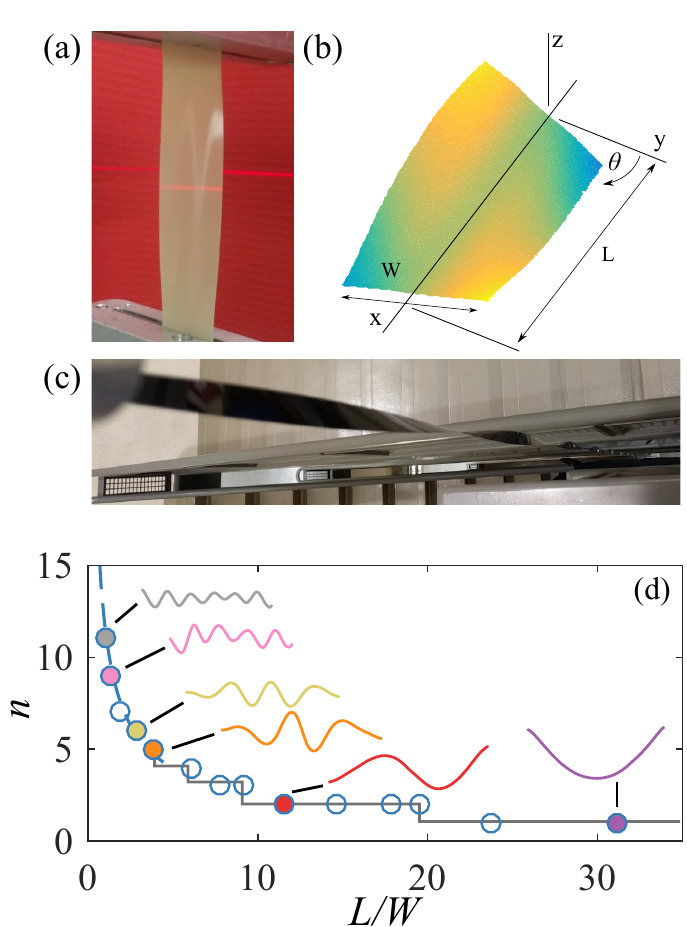}
\end{center}
\caption{(a) An image of wrinkled sheet which is clamped at its ends and twisted about its axis. (PolyVinyl sheet with $W = 30$\,mm and $t = 230\,\mu$m.) The intersection of a planar laser sheet with the wrinkled sheet is used to obtain the deflection of the ribbon surface. 
(b) A 3D reconstruction of a wrinkled  sheet obtained by sweeping the laser sheet ($\theta = 90^{\circ}$, $T = 0.1$, $L/W = 3$, $t/W = 0.0025$). (c) Image of a twisted ribbon in the stairwell of the Math-Physics Building ($L = 16$\,m). (d) Observed number of modes decreases to one as the sheet length to width ratio is increased. Examples of observed transects at various $L/W$ are also shown ($t/W = 3 \times 10^{-3}$). (Latex sheet with $W = 50$\,mm and $t = 152\,\mu$m.)
}
\label{fig1}
\end{figure}

At the other extreme, in the ribbon limit ($L/W \gg 1$), much more is known about the distribution of stress in the helicoid base state~\cite{Chopin2013,Chopin2015}. With respect to stability against buckling, two factors come into play. One is related to a tension-induced stiffness penalizing unstable modes with large amplitudes, and the other to a bending resistance which penalizes modes with large curvature. The tension-induced stiffness is analogous to that found in self-supported stretched sheets, and in thin films supported on elastic substrates~\cite{Cerda2003}. It is noteworthy that in the case of self-supported sheets, the tension-induced stiffness is not a mechanical constant but rather originates from nonlinear geometrical effects that dominate for large deflections~\cite{Cerda2003,Chopin2008,Davidovitch2011,Chopin2015}.. In particular, the tension-induced stiffness increases with the applied tension and decreases with sheet dimensions. Furthermore, tension-induced stiffness in twisted sheets is negligible compared to bending resistance when the tension is below a characteristic tension $T_C = (L^2t^2)/W^4$~\cite{Chopin2015}. 

Instead of tension, we find it more convenient here to introduce a new characteristic length scale, called the bendability length scale:
\begin{equation}
    L_B=(W^2/t)\sqrt{T}.
    \label{Eq:Linf}
\end{equation}
Thus, when $L \gg L_B$, tension-induced stiffness can be neglected. The bendability length is closely related to the more general concept of bendability number which has proven to be useful to address wrinkling instability in uniaxially stretched sheets~\cite{Davidovitch2011}. There, a high bendability number corresponds to a very thin sheet with negligible bending resistance compared to tension-induced stiffness.

Thus, the various possible regimes originally organized by Chopin, Demery, and Davidovitch~\cite{Chopin2015} in terms of tension $T$ can be recast in terms of $L$, 
$L_B$, and $L_{C}$ as follows.

(a) For very long lengths ($L\gg L_B$) or infinite length limit, the sheet buckles in the fundamental mode with a wavelength $\lambda \sim W$ when twisted above a critical twist angle which is L-independent. The transverse compression is balanced by the bending resistance alone, and that tension-induced stiffness can be considered to be negligible.     

(b) For intermediate lengths ($L_{C}\ll L \ll L_B$) or long but finite lengths, they calculated that the sheet buckles into higher modes (or wrinkles) with a wavelength which is smaller than $W$ at a  critical twist angle which decreases with tension and the length to thickness aspect ratio. For these lengths, the transverse compression has to overcome not only bending resistance but a tension-induced stiffness as well.

(c) For short lengths ($L\ll L_{C}$) where length and width of the sheets are comparable, the clamps can be important because the clamped boundary condition induces stretching which causes considerable deviations of the stress from that for a helicoidal base state. However, no predictions were available of this effect on the critical angle and wavelength.

In this paper, we discuss the transverse buckling instability in thin elastic sheets by measuring the critical instability angle and characterizing the wavelength of the buckling mode with twist over a wide range of sheet length, width, and thickness.  We find that the observed instabilities are consistent with the overall behavior proposed, with the critical twist angle dependent on the aspect ratios and the tension. However, some differences were also found. From our experiments, we identify three distinct transverse instabilities corresponding to three regimes with increasing sheet length while holding width constant. Namely, a clamp-dominated regime, a bendable regime, and a stiff regime. 

The clamp-dominated regime is typically reached for short sheet $L \lesssim L_C$. We demonstrate that the sheet destabilizes into higher buckling modes at relatively higher twist and shorter wavelength. We argue that the clamped boundary conditions at the ends inhibits out-of-plane deflection leading to a delayed buckling instability. In the bendable regime reached for intermediate length $L_C \lesssim L \lesssim L_B$, fundamental and higher-order buckling modes are observed above a L-dependent critical twist angle which scales as $\eta_B\sim \sqrt{t/L}{T}^{-1/4}$. When higher-order buckling develops, we find the wavelength $\lambda_B \sim \sqrt{tL}T^{-1/4}$. These scalings are consistent with theoretical predictions derived in the bendable regime~\cite{Chopin2015}. However, quite surprisingly, we find in the bendable regime that the sheet destabilizes into the fundamental buckling mode over a significant range of length to width aspect ratio. In this case, the wavelength trivially scales as  $\lambda_{tr} \sim W$, a feature that was not predicted by the theory. Finally, the stiff regime is identified by going to extreme lengths. The sheet is found to destabilize into the fundamental mode above a critical twist that scales as $\eta_S \sim t/W T^{-1/2}$ in full agreement with predictions in the long length limit. Thus, we find that the crossover length $L_B$ between the bendable regime and the stiff regime is captured by the transition from length-dependent to length-independent critical twist, but not by the wavelength transition from higher modes to the fundamental buckling mode. We find that the crossover length between the L-dependent to the L-independent critical twist is well captured by $L_B$ but the transition from higher modes to fundamental modes as $L$ is increased is significantly overestimated. 

\section{Experimental System} 

\begin{table}[t]
\centering
\begin{tabular}{cccccc}
\hline
Material & $E$ (MPa) & $\nu$ & $L$ (m) & $W$ (mm) & $t$ ($\mu$m)\\
\hline
PolyVinyl  & 1.2 & 0.38 & 0.05\,-\,0.2\,\,\, & 30, 50\,\,\, & 230, 500\\ 
Latex  & 3.2 & 0.50 & 0.08\,-\,2.0\,\,\, & 50, 80\,\,\, &  152, 500\\ 
PET & 5700 & 0.37 & 0.10\,-\,16.0 & 0.38, 12.7 & 10, 18\\ 
\hline
\end{tabular}
\caption{List of various materials used in the experiments and their properties.}
\label{tab:prop}
\end{table}

An image of a wrinkled sheet along with a 3D rendering and the coordinate system is shown in Fig.~\ref{fig1}(a,\,b), respectively. The experiments were performed in the laboratory for lengths less than $L =2$\,m, about the height of the lab. The stairwell in the physics department was used for longer lengths as shown in Fig.~\ref{fig1}(c).  As in our previous study~\cite{Chopin2013}, the ribbon is held under clamped boundary conditions at two opposite sides and twisted about its symmetry axis. The sheet is twisted by an angle $\theta$ about the $x$-axis while being pulled at the two clamped ends with a constant force $F$ which is applied along the $x$-axis with the help of linear guides. The materials used and their properties are listed in Table~\ref{tab:prop}. Then, the nondimensional tension $T = F/(E t W)$, where $E$ is the Young's modulus, and the normalized twist angle $\eta = \theta (W/L)$. In the experiments discussed in the following, we apply a tension $T > T_\lambda$, corresponding to the tension below which compressive stresses develop in the longitudinal direction that can give rise to longitudinal wrinkles~\cite{Green1937,Chopin2013}. 

Laser profilometry is used to obtain the shape of the ribbon. In this technique~\cite{Blair2005}, the sheet is illuminated with a laser and a cylindrical lens, resulting in an illumination pattern  which is proportional to out of plane deflection as shown in Fig.~\ref{fig1}(a). After sweeping the laser across the sheet surface, and calibrating for the viewing angle, we obtain the surface profile of the wrinkled sheet as shown in Fig.~\ref{fig1}(b) along with the coordinate system.

The amplitude of the deflection $\xi(x,y)$ from the $x-y$ plane is shown superimposed on the 3D rendering of the sheet. One observes that the largest number of wrinkles and amplitudes occur at the midsection of the sheet, and decay smoothly to zero toward the clamped edges. This occurs because of the boundaries conditions at $x=0$, where $\xi=0$ and $\partial \xi/\partial y = 0 $ for $-W/2 \le y \le W/2$, and  at $x=L$, where $\xi= y\tan\theta $ and $\partial \xi/\partial y = 0 $ for $-W/2 \le y \le W/2$. Because we find the maximum deflection and the wrinkles occur along the central transect, we focus in the following on the profile observed in this crosssection to identify the onset of the instability and the mode number.

\section{Aspect ratio dependent instabilities}
  
Fig.~\ref{fig1}(d) shows the mode number $n$ as a function of sheet length to width aspect ratio $L/W$ for a thin latex sheet ($t/W = 0.003$, $T = 0.08$). Example transects obtained using the laser profilometry  above the onset of transverse instability are also shown for several $L/W$ ratios. Because no moment is applied at the free edges, the transect appears to be curvature free at the edges. The mode number is then identified from the number of antinodes observed in the transect where the curvature passes through a maxima or minima. Following the plot from right to left, one observes that $n$ increases from $n=1$ (fundamental buckling mode or buckle)  to $n=12$ (higher-order buckling mode or wrinkle) as $L/W$ is decreased to 1 in this example.

Now, using Eq.~\ref{Eq:Lc} and Eq.~\ref{Eq:Linf} and substituting the material parameters corresponding to the elastic sheet used, we find that $L_C/W = 4$ and $L_B/W = 100$. Recalling that for $L < L_B$, the theory predicts that the ribbon wrinkles, i.e. higher-order buckling modes grows, it is worthwhile noting that the crossover between fundamental and higher-order buckling modes occurs for significantly smaller aspect ratio $L/W \approx 20$ according to our data than expected by the theory. However, it is unclear at the moment if the apparent discrepancy is due to a large numerical prefactor of order 10 in the scaling law, or due to a deeper issue with the derivation of Eq.~\ref{Eq:Linf}.

\subsection{Length-dependent instabilities}

\begin{figure}
\centering
\includegraphics[width = 8.5cm]{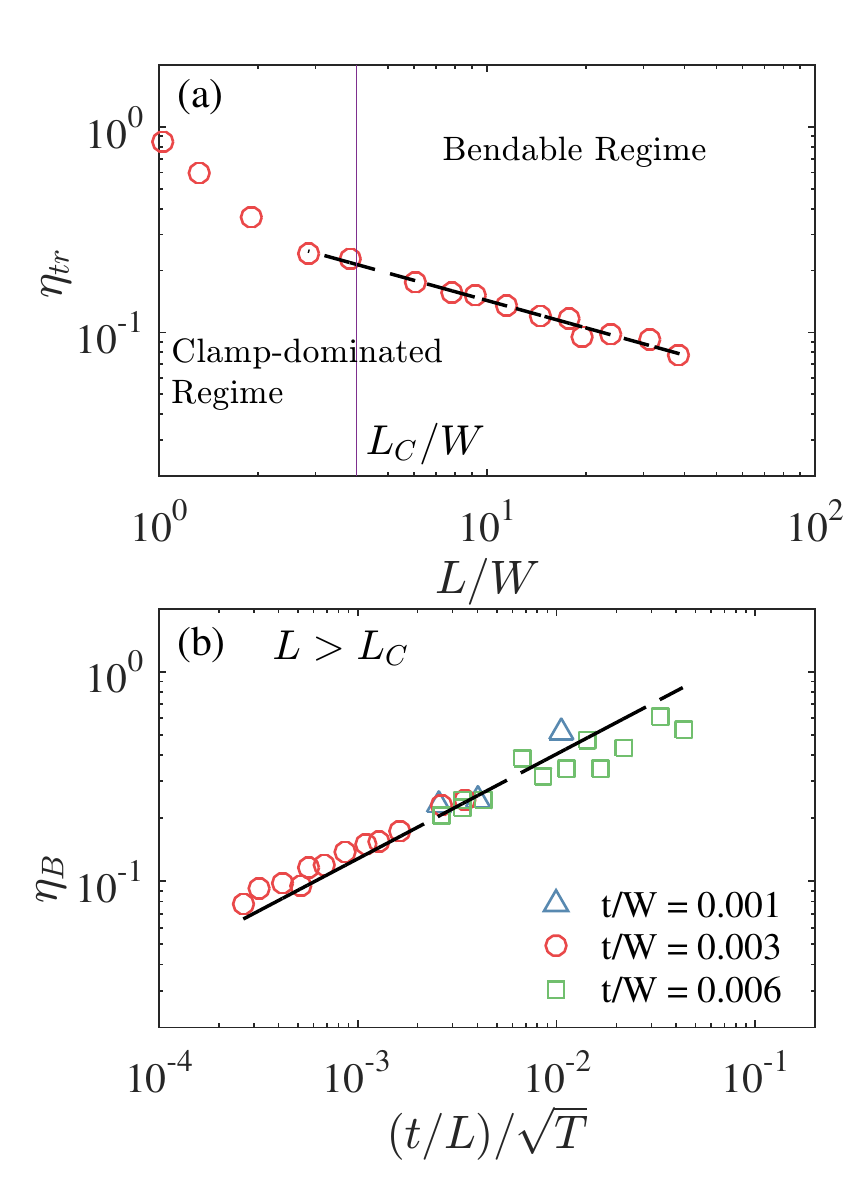}
\caption{(a) Measured $\eta_{tr}$ decreases as a function of $L/W$ with slope consistent with $1/2$ for $2 <L/W < 20$. The vertical line separates the clamp-dominated regime and the bendable regime. (b) The measured critical angle in the bendable regime $\eta_B$ as a function of $(t/L)/\sqrt{T} $ corresponding to $L/W > 2$ collapses on to a line with slope 1/2.}
\label{fig2}
\end{figure}

We obtain the critical twist $\eta_{tr}$ when a buckling mode starts to grow, by applying a prescribed tension and then slowly increasing the twist while monitoring the sheet deflection along the mid-transect of the sheet. Fig.~\ref{fig2}(a)  shows the measured $\eta_{tr}$ versus $L/W$ plotted in  log-log scale using the  same experimental conditions as in Fig.~\ref{fig1}(d). We observe that $\eta_{tr}$ decreases rapidly at first, before decreasing more steadily with $L/W$. We focus first on $\eta_{tr}$ for relatively large length for which precise predictions are available. In the bendable regime ($L_C\ll L \ll L_B$), the helicoid base state becomes unstable against higher-order buckling modes above a critical twist that scales as~\cite{Chopin2015} 
\begin{equation}
\eta_B = \alpha_{B} \sqrt{t/L} T^{-1/4},
\label{eq:etaT}
\end{equation}
for $\eta_{tr} \ll 1$ and $\lambda_{tr} \ll W$. Here, $\alpha_B$ is a numerical prefactor which is yet to be calculated, but can be determined from our data. The scaling is obtained from a linear stability analysis assuming a stretched helicoid base state. (Our previous experimental measurements of the ribbon morphology showed that this assumption is valid except very near the clamped edges~\cite{Chopin2013}.) This scaling corresponds to a line with slope $1/2$ in the case where $T$ and $t$ are held constant, and is shown along with the data in Fig.~\ref{fig2}(a). We find that the observed $\eta_{tr}$ is well aligned with this prediction for $L/W > 2$.  Interestingly, no change of scaling is observed at $L/W \approx 20$ when the instability reaches the fundamental mode $n=1$. 

\begin{figure}
\centering
\includegraphics[width = 8.5cm]{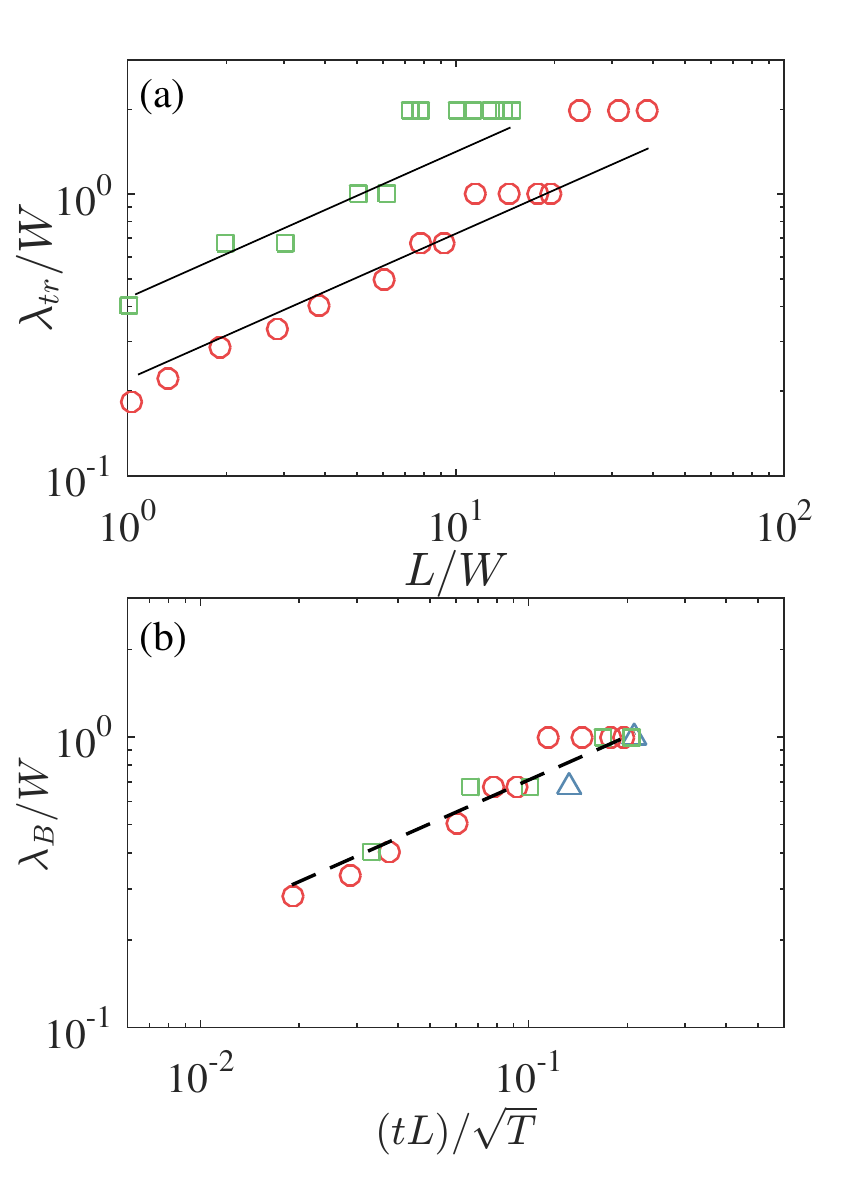}
\caption{(a) The measured wavelength $\lambda_{tr}$ as a function of $L/W$ is observed to increase till the fundamental mode is reached where $\lambda_{tr} = 2W$.  (b) The measured wavelength $\lambda_{tr}$ as a function of $(tL)/\sqrt{T}$ collapses on to a line with slope 1/2 in the wrinkling regime when $\lambda < W$ consistent with predictions in the bendable regime.}
\label{fig3}
\end{figure}

We measure $\eta_{tr}$ to test Eq.~\ref{eq:etaT} in the bendable regime more extensively over a wide range of applied tension, sheet thickness, and materials listed in Table~\ref{tab:prop}. The observed $\eta_{tr}$ in the $L$-dependent regime where $ L < L_{B}$, but above the point where edge effects start to dominate ($L>L_C$), is plotted in Fig.~\ref{fig2}(b) as a function of $(t/L)/\sqrt{T}$.  We observe excellent agreement with the predicted scaling and find $\alpha_B = 4.0 \pm 0.3$. 

Now, the corresponding wavelength of the wrinkles in this bendable regime is given by~\cite{Chopin2015}
\begin{equation}
\lambda_{B} = \alpha_{\lambda} \sqrt{L t} T^{-1/4},
\label{eq:lambda}
\end{equation} 
where $\alpha_{\lambda}$ is a numerical prefactor. Here, the scaling with tension can be noted to be the same as for longitudinal wrinkles~\cite{Chopin2013}.  The difference is the dependence here on sheet length $L$ rather than the width $W$ in the case of the longitudinal wrinkles~\cite{Chopin2013}.    

The wavelength $\lambda_{tr}$ obtained as $2W/n$ is plotted in Fig.~\ref{fig3}(a) as a function of length and observed to increase like a staircase function till the maximum wavelength corresponding to twice the width of the sheet is reached. The data is observed to be well aligned with scaling given in Eq.~\ref{eq:lambda} for $\lambda_{tr}/W \ll 1$, with systematic deviations growing as $\lambda_{tr}/W$ approaches 1. This measured trend is consistent with the estimate calculated in Ref.~\cite{Chopin2015} with systematically higher wavelengths for the thicker ribbon. Now, plotting  $\lambda_{tr}$ versus $(L t)/\sqrt{T}$ for the latex as well as the PET sheets, we again observe good collapse of the data onto a line of slope $1/2$ expected from the theory, provided $\lambda_{tr} < W$.   We find that $\alpha_{\lambda} \approx 2.2 \pm 0.1$. It is noteworthy that the material parameters have been varied over three orders of magnitude in obtaining this data and gives a sense of the robustness of the scaling and the determined $\alpha_{\lambda}$.

Thus, we find consistency with the prediction that wrinkling occurs in twisted sheets which depends on the applied tension in the limit of thin ribbons. This transverse instability occurs at lower twist angle with increasing tension. This is opposite even qualitatively to the trend at low tension where longitudinal wrinkling occurs~\cite{Green1937,Chopin2013}. In that case, the instability occurs at increasing twist angle as the tension is increased until the crossover tension $T_\lambda$ is reached. 

Furthermore, the points corresponding to the thicker latex ribbon ($t/W = 0.006$ and $L_{B} = 30$) can be noted to be especially interesting and may point to a larger range of validity for the scaling shown in Eq.~\ref{eq:etaT} than implied by the calculation assuming $\lambda_{tr} < W$. In particular it can be noted that for this thicker ribbon, $\eta_{tr}$ is observed to scale consistent with Eq.~\ref{eq:etaT} even though the fundamental mode is observed over a considerable part of this range. Thus, the scaling appears tied more strongly to the length dependence of $\eta_{tr}$ rather than the condition that $\lambda_{tr} \ll W$. Further theoretical developments are still necessary to better understand this regime. 

Focusing on the small $L/W$ limit in Fig.~\ref{fig2}(a), where $ L \sim  L_C$, the trend in the data shows that elastic sheets become unstable and develop higher-order buckling modes above a critical threshold $\eta_{tr}$ which is found to be significantly larger than the predictions given by Eq.~\ref{eq:etaT}. Further, the measured $\lambda_{tr}$ is found to be slightly smaller than predicted by Eq.~\ref{eq:lambda} in the same range of $L/W$. Here, we argue that edge effects are responsible for the significant deviations of the measured threshold and wavelength from predictions.  We note that the prediction $L_{C}/W = 4$ is consistent with a change in scaling for $\eta_{tr}$ (see dashed in Fig.~\ref{fig2}(a)). This result suggests that the clamped edges are responsible for delaying the appearance of wrinkles for $L/W \gtrsim 1$, but do not suppress the instability. 

The observed wrinkles in the clamp-dominated regime bear some similarity to tensional wrinkles observed at the center of uniaxially stretched sheets~\cite{friedl2000buckling,Cerda2003,healey2013wrinkling}. In that configuration, the uniaxial state of stress of a stretched membrane is frustrated by the clamped boundaries which induce shear and transverse stresses~\cite{nayyar2011stretch}. Therefore, development of transverse compressive stresses can give rise to an instability driven by the clamped edge stresses~\cite{Cerda2003}. Later numerical studies indicate that the wrinkling instability  in fact occupies only a bounded region of the $L/W$-$T$ phase diagram~\cite{healey2013wrinkling}. However, in spite of these developments, the fundamental reason for the development of a compressive zone away from the boundary remains unclear. 

Now considering our twisted sheet configuration, we also argue that the frustration of the helicoid base state by the clamped edges  is an essential ingredient to explain the delayed wrinkling instability. However, instead of giving rise to the wrinkling mechanism in axially stretched sheets, the clamped edges of a twisted sheet appear to act as a stabilizing effect in determining $\eta_{tr}$. We reach this conclusion because of the relatively higher rise in $\eta_{tr}$ in the clamp-dominated regime compared to the bendable regime seen in Fig.~\ref{fig2}(a), and relatively lower wavelengths as well in Fig.~\ref{fig3}(a). A theoretical approach of the wrinkling mechanism in this regime which includes both twist and stretch loading at small $L/W$ is not available, and is outside the scope of this study.

\subsection{Length-independent Instability}

In the limit where $\eta_{tr}$ becomes independent of $L$, Chopin, Demery, and Davidovitch~\cite{Chopin2015} calculated that a novel buckling regime would be reached where the sheet destabilizes in the fundamental $n=1$ mode at a critical twist angle in the stiff regime 
\begin{equation}
\eta_{S} = \alpha_{S} (t/W) T^{-1/2}, 
\label{eq:eta-long}
\end{equation}
where $\alpha_{S} = 4.4$ was obtained numerically at large $T$. For small $T (< T_{\lambda})$ and in the limit $\eta^2/T \gg 1$, the development of the longitudinal wrinkling instability far-from-threshold allows one to approximate the ribbon base state as being essentially a helicoid stretched in the vicinity of the free edges with a vanishing compression everywhere else. Using a linear stability analysis with reference to this post-buckling base state, $\alpha_{S} = \pi/\sqrt{3}$ was calculated analytically.  

\begin{figure}
\centering
\includegraphics[width = 8.5cm]{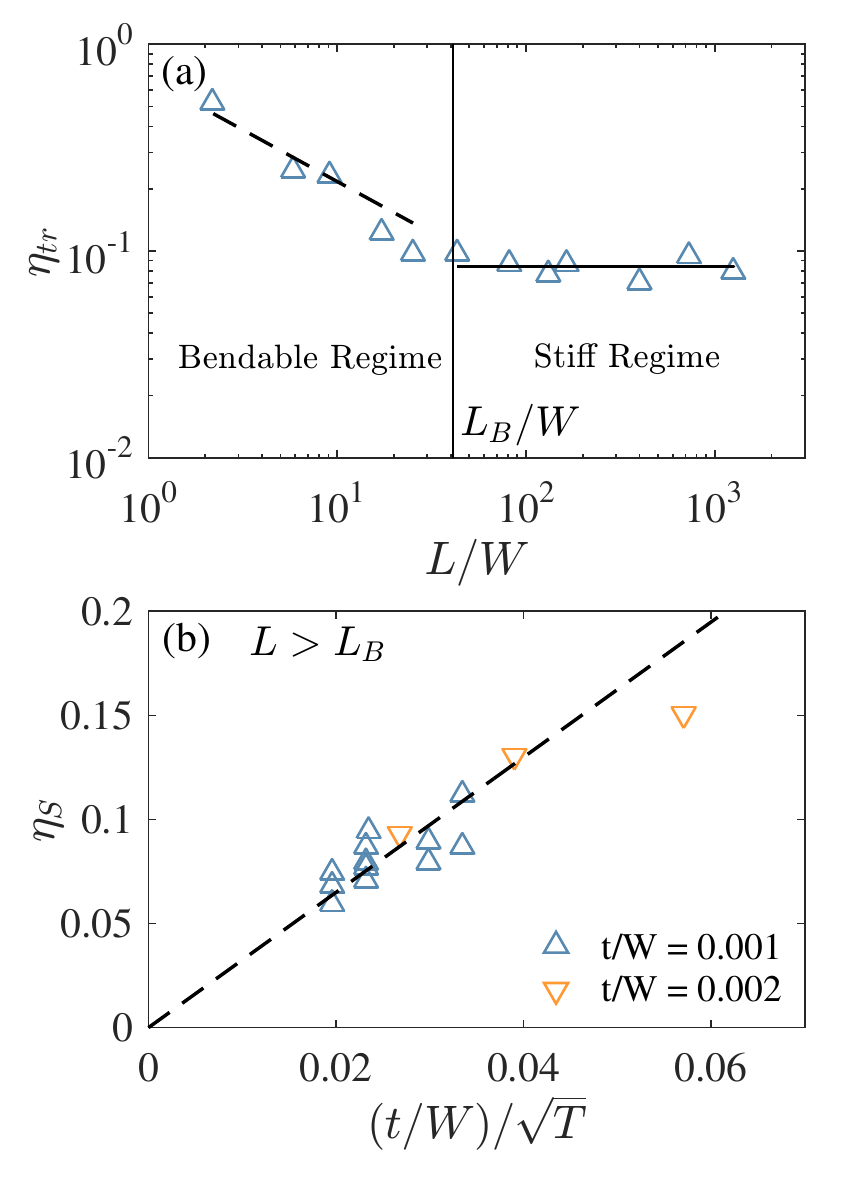}
\caption{(a)  $\eta_{tr}$ crosses over from decreasing inversely as the length  to becoming  $L$-independent as the length is increased for the PET ribbon held at constant tension $T = 0.003$. The crossover occurs as the contribution of the L-dependent tensional stiffening to the stability decreases to zero relative to the L-independent bending contribution. (b) The critical twist angle in the stiff regime $\eta_{S}$ plotted as a function of $(t/W)/\sqrt{T}$ along with linear fit given by Eq.~\ref{eq:eta-long}.}
\label{fig4}
\end{figure}

To reach this regime, we now consider extremely long ribbon experiments performed in the stairwell (see Fig.~\ref{fig1}(c)), in addition to those performed in the lab. The measured $\eta_{tr}$ as a function of $L/W$ for $L > 2$\,m is shown in Fig.~\ref{fig4}(a). For $L < L_B$, we observe scaling consistent with Eq.~\ref{eq:etaT}, but then as $L$ is increased above $L_B$, clear deviations are observed as $\eta_{tr}$ occurs at a constant value within the error of measurements which is approximately $\pm 5$\% in this case.

To understand the effect of this length independence on the scalings, $\eta_{tr} \equiv \eta_S$ is plotted in a linear-linear scale as a function of $(t/W)/\sqrt{T}$ in Fig.~\ref{fig4}(b) for the data corresponding to the $L$-independent regime. A linear fit according to Eq.~\ref{eq:eta-long} with $\alpha_T = 3.2 \pm 0.2$ is also shown which is consistent with previous calculations. However, significant deviations can be also noted from this form which are somewhat higher than the error in the identification of $\eta_{tr}$. In the case of the PET ribbons used here, lowering the $T$ resulted in approaching $T_\lambda$, the transition to the longitudinal limit, while increasing $T$ beyond the reported range resulted in plastic deformation. Further experiments are needed to fully test this scaling over a wider range of $T$. However, this is beyond the scope of the materials available to us.  Nonetheless, it is clear from Fig~\ref{fig4}(a) that the instability occurs at a much higher twist than predicted by Eq.~\ref{eq:etaT}, clearly demonstrating that the nature of the instabilities in the bendable and stiff regime are different.

To understand the two different mechanisms operating in the L-dependent bendable regime, and the L-independent stiff regime, we start by identifying the forces acting normal to the sheet and examining their relative contributions. Besides the transverse compressive force which is driving the instability, two stabilizing forces act on the sheet: a bending resistance penalizing large curvature, (or equivalently, small wavelength), and a tension-induced restoring force which prevents the development of large amplitudes. For large $L/W$, the tension-induced force is not sufficiently high to penalize the fundamental mode in favor of higher-order modes of smaller amplitude but larger curvature, thus the fundamental buckling mode is observed. As $L/W$ is decreased, the tension induced force is of the same order as the bending resistance, which indicates that wrinkling modes start to be energetically favorable. This crossover length between a L-dependent to a L-independent critical twist is observed to coincide well with the predicted $L_B$. By contrast, the transition from mode number $n=1$ to $n>1$ is observed to occur at a significantly smaller length than $L_B$.

\section{Conclusions}

We have experimentally studied the transverse wrinkling of a thin elastic sheet held under tension and twisted about its long axis.  The critical twist is found to be not only dependent on the aspect ratio of the sheet, but also on the applied tension along the axis around which the sheet is twisted. Three distinct regimes are identified, consistent with recent theoretical model of transverse buckling developed starting from a covariant form of the FvK equations.

To rationalize the different instability regimes, we introduced two characteristic lengthscales: a clamp length $L_C$ and a bendability length $L_B$. When $L > L_B$, the instability is $L$-independent. We find that the sheet destabilizes in the fundamental buckling mode and that the critical twist  decreases proportional to the thickness, and as inverse of the width and the square root of the applied tension. This instability occurs in the stiff regime as the cross-section is only slightly curved. When the length is decreased below $L_B$, the instability becomes $L$-dependent. For intermediate length $L > L_C$, the sheet destabilizes into fundamental or higher-order buckling modes. We identify these instabilities with the bendable regime. When higher-order buckling modes develop,  the critical twist and the wavelength of the wrinkles slowly but systematically decrease as the fourth-root of tension. However, it was unanticipated that the fundamental mode can develop in the bendable regime characterized by a $L$-dependent threshold. At even smaller length $L<L_C$ in the clamped regime, the clamped boundary conditions delays the development of the instability with greater twist required to wrinkle the sheet because the sheet is under tension near the boundaries along the transverse direction. 

Thus, our experiments provide a thorough test of the scaling approach and the regimes of their applicability, as opposed to direct numerical simulations of the thin plate equations which, while accurate, give rise to less insight on the development of the instabilities, and the various operative mechanisms. This approach also yields simple forms for the dependence on material parameters.  Our study provides the prefactors in addition to testing the derived scaling laws against materials with Young's modulus distributed over three orders of magnitude.   

\section*{Acknowledgments}
We thank A. Panaitescu and M. Hannout for help with experiments, C. Trimble for preliminary work, and B. Davidovitch for stimulating discussions. This work was supported by the National Science Foundation under grant number DMR 1508186.


\begin{thebibliography}{99}

\bibitem{Coulomb1784}
Coulomb CA. 1784  Recherches th{\'e}oriques et exp{\'e}rimentales sur la force
  de torsion. {\em M{\'e}moire de l'Acad{\'e}mie des Sciences Paris}
  \textbf{66}, 229--272.

\bibitem{SaintVenant1855}
de~Saint-Venant A. 1855 {\em De la torsion des prismes}.
Imprimerie Imp{\'e}riale.

\bibitem{love2013treatise}
Love AEH. 2013 {\em A treatise on the mathematical theory of elasticity}.
Cambridge university press.

\bibitem{timoshenko1953history}
Timoshenko S. 1953 {\em History of strength of materials: with a brief account
  of the history of theory of elasticity and theory of structures}.
Dover Publications.

\bibitem{antman1981large}
Antman SS, Kenney CS. 1981  Large buckled states of nonlinearly elastic rods
  under torsion, thrust, and gravity. {\em Archive for rational mechanics and
  analysis} \textbf{76}, 289--338.

\bibitem{thompson1996helix}
Thompson J, Champneys A. 1996  From helix to localized writhing in the
  torsional post-buckling of elastic rods. {\em Proc. R. Soc. Lond. A}
  \textbf{452}, 117--138.

\bibitem{Goriely2001}
Goriely A, Nizette M, Tabor M. 2001  On the dynamics of elastic strips. {\em J.
  Nonlinear Sci.} \textbf{11}, 3--45.

\bibitem{van1998lock}
Van~der Heijden G, Thompson J. 1998  Lock-on to tape-like behaviour in the
  torsional buckling of anisotropic rods. {\em Physica D} \textbf{112},
  201--224.

\bibitem{van2003instability}
Van~der Heijden G, Neukirch S, Goss V, Thompson J. 2003  Instability and
  self-contact phenomena in the writhing of clamped rods. {\em Int. J. Mech.
  Sci.} \textbf{45}, 161--196.

\bibitem{Dias2015}
Dias MA, Audoly B. 2015  "Wunderlich, Meet Kirchhoff": A General and Unified
  Description of Elastic Ribbons and Thin Rods. {\em J. Elast.} \textbf{119},
  49--66.

\bibitem{chopin2016ordered}
Chopin J, Romildo~Filho T. 2016  Ordered Crumpled States in Twisted Ribbons.
  {\em arXiv preprint arXiv:1603.02081}.

\bibitem{Audoly2010}
Audoly B, Pomeau Y. 2010 {\em Elasticity and geometry}.
Oxford University Press.

\bibitem{Cerda2003}
Cerda E, Mahadevan L. 2003  Geometry and physics of wrinkling. {\em Phys. Rev.
  Lett.} \textbf{90}, 074302.

\bibitem{Chopin2008}
Chopin J, Vella D, Boudaoud A. 2008  The liquid blister test. {\em Proc. R.
  Soc. Lond. A} \textbf{464}, 2887--2906.

\bibitem{Vella2009Macroscopic}
Vella D, Bico J, Boudaoud A, Roman B, Reis PM. 2009  The macroscopic
  delamination of thin films from elastic substrates. {\em Proc. Natl. Acad. of
  Sci. U.S.A.} \textbf{106}, 10901--10906.

\bibitem{Vandeparre2011}
Vandeparre H, Pineirua M, Brau F, Roman B, Bico J, Gay C, Bao WZ, Lau CN, Reis
  PM, Damman P. 2011  Wrinkling Hierarchy in Constrained Thin Sheets from
  Suspended Graphene to Curtains. {\em Physical Review Letters} \textbf{106},
  224301.

\bibitem{brau2011multiple}
Brau F, Vandeparre H, Sabbah A, Poulard C, Boudaoud A, Damman P. 2011
  Multiple-length-scale elastic instability mimics parametric resonance of
  nonlinear oscillators. {\em Nat. Phys.} \textbf{7}, 56.

\bibitem{Davidovitch2011}
Davidovitch B, Schroll RD, Vella D, Adda-Bedia M, Cerda EA. 2011  Prototypical
  model for tensional wrinkling in thin sheets. {\em Proc. Natl. Acad. of Sci.
  U.S.A.} \textbf{108}, 18227--18232.

\bibitem{chopin2017dynamic}
Chopin J, Dasgupta M, Kudrolli A. 2017  Dynamic Wrinkling and Strengthening of
  an Elastic Filament in a Viscous Fluid. {\em Phys. Rev. Lett.} \textbf{119},
  088001.

\bibitem{Mockensturm2001}
Mockensturm EM. 2001  The elastic stability of twisted plates. {\em J. Appl.
  Mech.} \textbf{68}, 561--567.

\bibitem{Chopin2013}
Chopin J, Kudrolli A. 2013  Helicoids, Wrinkles, and Loops in Twisted Ribbons.
  {\em Phys. Rev. Lett.} \textbf{111}, 174302.

\bibitem{Green1937}
Green AE. 1937  The elastic stability of a thin twisted strip - II. {\em Proc.
  R. Soc. A} \textbf{161}, 197--220.

\bibitem{Coman2008}
Coman CD, Bassom AP. 2008  An asymptotic description of the elastic instability
  of twisted thin elastic plates. {\em Acta Mech.} \textbf{200}, 59--68.

\bibitem{Chopin2015}
Chopin J, Demery V, Davidovitch B. 2015  Roadmap to the Morphological
  Instabilities of a Stretched Twisted Ribbon. {\em J. Elast.} \textbf{119},
  137--189.

\bibitem{chopin2016disclinations}
Chopin J, Kudrolli A. 2016  Disclinations, e-cones, and their interactions in
  extensible sheets. {\em Soft Matter} \textbf{12}, 4457--4462.

\bibitem{Korte2011}
Korte AP, Starostin EL, van~der Heijden GHM. 2011  Triangular buckling patterns
  of twisted inextensible strips. {\em Proc. R. Soc. A} \textbf{467}, 285--303.

\bibitem{Kit2012}
Kit OO, Tallinen T, Mahadevan L, Timonen J, Koskinen P. 2012  Twisting graphene
  nanoribbons into carbon nanotubes. {\em Phys. Rev. B} \textbf{85}, 085428.

\bibitem{Blair2005}
Blair DL, Kudrolli A. 2005  Geometry of crumpled paper. {\em Phys. Rev. Lett.}
  \textbf{94}, 166107.

\bibitem{friedl2000buckling}
Friedl N, Rammerstorfer FG, Fischer FD. 2000  Buckling of stretched strips.
  {\em Computers \& structures} \textbf{78}, 185--190.

\bibitem{healey2013wrinkling}
Healey TJ, Li Q, Cheng RB. 2013  Wrinkling behavior of highly stretched
  rectangular elastic films via parametric global bifurcation. {\em J.
  Nonlinear Sci.} \textbf{23}, 777--805.

\bibitem{nayyar2011stretch}
Nayyar V, Ravi-Chandar K, Huang R. 2011  Stretch-induced stress patterns and
  wrinkles in hyperelastic thin sheets. {\em Int. J. Solids Struct.}
  \textbf{48}, 3471--3483.

\end{thebibliography}
\end{document}